# Online $^{222}$Rn removal by cryogenic distillation in the XENON100 experiment

E. Aprile[1], J. Aalbers[2], F. Agostini[3,4], M. Alfonsi[5], F. D. Amaro[6], M. Anthony[1], F. Arneodo[7], P. Barrow[8], L. Baudis[8], B. Bauermeister[5,9], M. L. Benabderrahmane[7], T. Berger[10], P. A. Breur[2], A. Brown[2], E. Brown[10], S. Bruenner[11,a], G. Bruno[3], R. Budnik[12], L. Bütikofer[13,b], J. Calvén[9], J. M. R. Cardoso[6], M. Cervantes[14], D. Cichon[11], D. Coderre[13,b], A. P. Colijn[2], J. Conrad[9,c], J. P. Cussonneau[15], M. P. Decowski[2], P. de Perio[1], P. Di Gangi[4], A. Di Giovanni[7], S. Diglio[15], E. Duchovni[12], G. Eurin[11], J. Fei[16], A. D. Ferella[9], A. Fieguth[17], D. Franco[8], W. Fulgione[3,18], A. Gallo Rosso[3], M. Galloway[8], F. Gao[1], M. Garbini[4], C. Geis[5], L. W. Goetzke[1], L. Grandi[19], Z. Greene[1], C. Grignon[5], C. Hasterok[11], E. Hogenbirk[2], R. Itay[12], B. Kaminsky[13,b], G. Kessler[8], A. Kish[8], H. Landsman[12], R. F. Lang[14], D. Lellouch[12], L. Levinson[12], M. Le Calloch[15], Q. Lin[1], S. Lindemann[11,13], M. Lindner[11], J. A. M. Lopes[6,d], A. Manfredini[12], I. Maris[7], T. Marrodán Undagoitia[11], J. Masbou[15], F. V. Massoli[4], D. Masson[14], D. Mayani[8], Y. Meng[20], M. Messina[1], K. Micheneau[15], B. Miguez[18], A. Molinario[3], M. Murra[17,e], J. Naganoma[21], K. Ni[16], U. Oberlack[5], S. E. A. Orrigo[6,f], P. Pakarha[8], B. Pelssers[9], R. Persiani[15], F. Piastra[8], J. Pienaar[14], M.-C. Piro[10], V. Pizzella[11], G. Plante[1], N. Priel[12], L. Rauch[11], S. Reichard[14], C. Reuter[14], A. Rizzo[1], S. Rosendahl[17,g], N. Rupp[11], R. Saldanha[19], J. M. F. dos Santos[6], G. Sartorelli[4], M. Scheibelhut[5], S. Schindler[5], J. Schreiner[11], M. Schumann[13], L. Scotto Lavina[22], M. Selvi[4], P. Shagin[21], E. Shockley[19], M. Silva[6], H. Simgen[11], M. v. Sivers[13,b], A. Stein[20], D. Thers[15], A. Tiseni[2], G. Trinchero[18], C. Tunnell[2,19], N. Upole[19], H. Wang[20], Y. Wei[8], C. Weinheimer[17], J. Wulf[8], J. Ye[16], Y. Zhang[1], (XENON Collaboration)[h], and I. Cristescu[23]

[1] Physics Department, Columbia University, New York, NY, USA
[2] Nikhef and the University of Amsterdam, Science Park, Amsterdam, The Netherlands
[3] INFN-Laboratori Nazionali del Gran Sasso and Gran Sasso Science Institute, L'Aquila, Italy
[4] Department of Physics and Astrophysics, University of Bologna and INFN-Bologna, Bologna, Italy
[5] Institut für Physik & Exzellenzcluster PRISMA, Johannes Gutenberg-Universität Mainz, Mainz, Germany
[6] Department of Physics, University of Coimbra, Coimbra, Portugal
[7] New York University Abu Dhabi, Abu Dhabi, United Arab Emirates
[8] Physik-Institut, University of Zurich, Zurich, Switzerland
[9] Department of Physics, Oskar Klein Centre, Stockholm University, AlbaNova, Stockholm, Sweden
[10] Department of Physics, Applied Physics and Astronomy, Rensselaer Polytechnic Institute, Troy, NY, USA
[11] Max-Planck-Institut für Kernphysik, Heidelberg, Germany
[12] Department of Particle Physics and Astrophysics, Weizmann Institute of Science, Rehovot, Israel
[13] Physikalisches Institut, Universität Freiburg, 79104 Freiburg, Germany
[14] Department of Physics and Astronomy, Purdue University, West Lafayette, IN, USA
[15] SUBATECH, Ecole des Mines de Nantes, CNRS/In2p3, Université de Nantes, Nantes, France
[16] Department of Physics, University of California, San Diego, CA, USA
[17] Institut für Kernphysik, Westfälische Wilhelms-Universität Münster, Münster, Germany
[18] INFN-Torino and Osservatorio Astrofisico di Torino, Turin, Italy
[19] Department of Physics and Kavli Institute of Cosmological Physics, University of Chicago, Chicago, IL, USA
[20] Physics and Astronomy Department, University of California, Los Angeles, CA, USA
[21] Department of Physics and Astronomy, Rice University, Houston, TX, USA
[22] LPNHE, Université Pierre et Marie Curie, Université Paris Diderot, CNRS/IN2P3, Paris 75252, France
[23] Tritium Laboratory Karlsruhe, Karlsruhe Institute of Technology, Eggenstein-Leopoldshafen, Germany

[a] e-mail: stefan.bruenner@mpi-hd.mpg.de
[b] Also with Albert Einstein Center for Fundamental Physics, University of Bern, Bern, Switzerland
[c] Wallenberg Academy Fellow
[d] Also with Coimbra Engineering Institute, Coimbra, Portugal
[e] e-mail: michaelmurra@wwu.de

**Abstract** We describe the purification of xenon from traces of the radioactive noble gas radon using a cryogenic distillation column. The distillation column was integrated into the gas purification loop of the XENON100 detector for online radon removal. This enabled us to significantly reduce the constant $^{222}$Rn background originating from radon emanation. After inserting an auxiliary $^{222}$Rn emanation source in the gas loop, we determined a radon reduction factor of $R > 27$ (95% C.L.) for the distillation column by monitoring the $^{222}$Rn activity concentration inside the XENON100 detector.

## 1 Introduction

Intrinsic contaminations of radioisotopes in liquid xenon detectors are a serious background source in rare-event experiments such as searches for dark matter [1–4] and neutrinoless double beta-decay searches [5]. The two most important internal sources of radioactive backgrounds are $^{85}$Kr and $^{222}$Rn. While krypton can be removed by cryogenic distillation before the start of a measurement [6–9], $^{222}$Rn is continuously produced inside detector materials due to the decay of trace amounts of $^{226}$Ra. As a noble gas with a half-life of 3.8 days, $^{222}$Rn can enter the liquid xenon target by means of diffusion or recoil from prior $\alpha$-decays. Once in the liquid, $^{222}$Rn distributes homogeneously, reaching also the innermost part of the detector. Subsequent beta-decays in the $^{222}$Rn decay chain are sources of low energy background [2–4]. Active detector shielding and fiducialization, i.e., making use of the self-shielding properties of liquid xenon, cannot be used to reject them.

In order to reduce the $^{222}$Rn induced background, extensive screening campaigns are performed to select only materials with a low radon emanation rate [10,11]. An additional strategy is the development of a radon removal system (RRS) operated online. Integrated in a purification loop, contaminated xenon is circulated continuously from the detector through the RRS. Radon is retained in the RSS until disintegration, while purified xenon is flushed back into the detector. Since radon and xenon are both noble gases with similar physical properties, finding an adequate separation technique is challenging. A potential RRS based on adsorption is discussed in [12] for the XMASS liquid xenon detector. There, gaseous xenon is flushed through cooled charcoal traps where radon gets adsorbed more efficiently than xenon. However, a successful operation at the required low radon concentrations has not been reported yet. Independent studies demonstrate that the charcoal itself will emanate radon, limiting the reduction power of such an RRS [10].

Cryogenic distillation is an alternative separation technique, successfully used for removing krypton from xenon [6–9]. In contrast to krypton, radon is depleted in the boil-off gas above a liquid xenon reservoir due to its lower vapor pressure compared to xenon (single-stage distillation) [13]. This effect is enhanced in a multiple-stage distillation column which can be used as an RRS.

This paper describes the successful operation of a radon removal system based on cryogenic distillation for liquid xenon detectors. In a radon distillation campaign, we integrated a cryogenic distillation column [6] in the gas purification loop of the XENON100 experiment [9]. By monitoring the $^{222}$Rn activity concentration inside the detector, we investigated the capability of cryogenic distillation to reduce radon-induced background. In the distillation process, radon is trapped until disintegration instead of being removed by extraction of highly contaminated xenon as it is necessary for e.g. krypton removal. Thus, the RRS can be operated continuously with no loss of xenon.

The experimental setup and the expected $^{222}$Rn reduction, based on a rate equation model for our RRS, are presented in Sects. 2 and 3. Section 4 describes the selection of $^{222}$Rn events in XENON100 data, used to monitor the radon activity concentration during the operation. The observed radon reduction and the purification power of the used distillation column are presented in Sects. 5 and 6, respectively.

## 2 Experimental setup

The XENON100 detector [9] is equipped with a gas purification loop consisting of a gas transfer pump (pump 1), a mass-flow controller (MFC), and a high-temperature gas purifier (getter) to remove electronegative impurities as shown in Fig. 1. In order to integrate the distillation column, we extended the purification loop by an RRS gas interface. The radon-enriched xenon is flushed from the XENON100 detector via the RRS gas interface towards the distillation column (red line). A second gas transfer pump (pump 2) pushes the purified xenon back into the XENON100 detector, again via the RRS gas interface (blue line). The xenon flow is controlled by means of mass-flow controllers. The getter removes electronegative impurities before the xenon re-enters the detector.

An important element of the RRS gas interface is the auxiliary $^{222}$Rn emanation source. As we will discuss in Sect. 5, it has a constant $^{222}$Rn emanation rate of $(72 \pm 2)$ mBq. The source is placed directly before the distillation column and can be switched on/off with valves. In Sect. 6, we will make use of this source in order to determine the radon reduction capability of our RRS.

[f] Now at IFIC, CSIC-Universidad de Valencia, Valencia, Spain
[g] Now at Physikalisch-Technische Bundesanstalt (PTB), Braunschweig, Germany
[h] email: xenon@lngs.infn.it

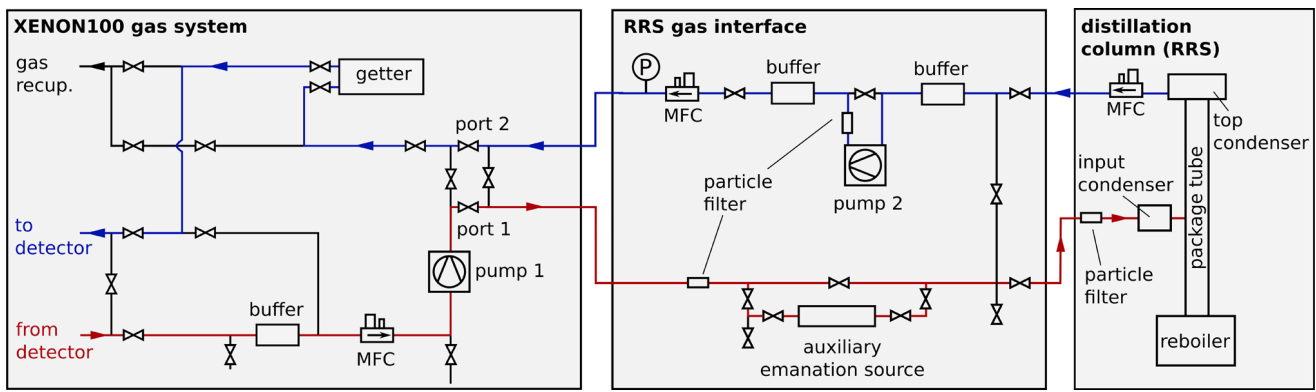

**Fig. 1** The xenon coming from the detector is guided from the XENON100 gas system (*left*) via the RRS gas interface (*middle*) to the distillation column (*right*), where radon is removed. The purified xenon is pumped back to the detector via the RRS gas interface. The purification flow is controlled via mass-flow controllers (MFC) in front of the two gas transfer pumps. A bypass to an auxiliary radon emanation source can be opened to artificially enhance the radon concentration in the detector

The distillation column was originally designed for the removal of krypton from xenon [6], but was used as a radon distillation column after a number of small modifications. It consists of four main components: the input condenser, the package tube, the top condenser and the reboiler which contains a liquid xenon reservoir of up to 8 kg. The inflowing xenon from the RRS gas interface is pre-cooled to $-96\,°C$ inside the input condenser and is then injected into the middle of the package tube of the column. There, a continuous counterflow is established between up-streaming xenon gas evaporated in the reboiler and down-streaming liquid xenon liquefied by the top condenser. The package tube has a height of 1.1 m and is filled with structured packing material which provides a large surface. This ensures that the up(down)-streaming xenon condenses (evaporates) in several distillation stages [6]. Due to its lower vapor pressure with respect to xenon, radon is enriched in the liquid phase in each distillation stage [13]. Thus, the liquid reservoir inside the reboiler becomes radon enriched, while gaseous xenon at the top of the column has the lowest radon concentration. Radon-depleted xenon is extracted and circulated back into the detector via a gas port close to the top condenser. The mass balance between gas inlet and outlet of the column is maintained by mass-flow controllers. In contrast to the standard distillation process used to remove krypton from xenon, radon distillation does not require the extraction of highly contaminated xenon (off gas), as radon stays inside the reboiler's liquid reservoir until disintegration. Thus, no xenon is lost during the online operation of the RRS.

## 3 Expected $^{222}$Rn reduction

In this section, we discuss the expected radon rate inside the XENON100 detector for a given reduction capability of the RRS [10]. Depending on their position in the detector and its gas system, we distinguish between two types of radon emanation sources: type 1 sources are located inside the XENON100 cryostat or after the RRS within the gas purification loop. Thus, radon emanated from these sources reaches the detector first before it is circulated through the RRS. According to Fig. 1, obvious type 1 sources in our setup are the XENON100 detector itself, pump 2 in the RRS gas interface and the getter in the XENON100 gas system. A radon source is referred to as type 2, if it is located in the purification system between the detector and the RRS (e.g., the pump 1 in the XENON100 gas system). Figure 2 shows a simplified diagram of the different contributions to the final radon activity in XENON100.

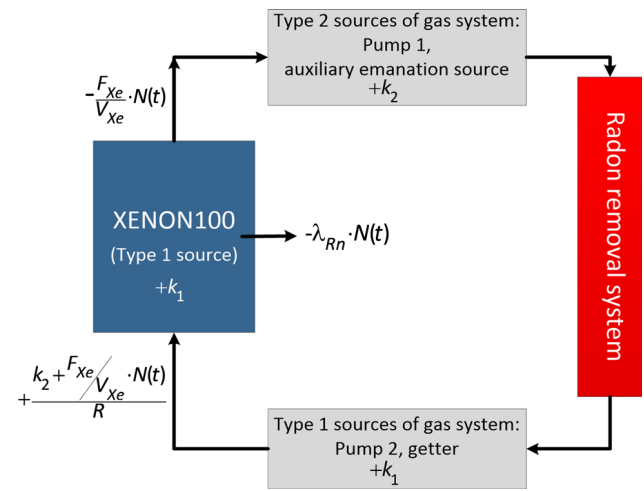

**Fig. 2** Diagram of the different radon sources contributing to the radon activity in the detector: the experimental setup can be divided into the XENON100 detector, the radon removal system and the gas system before/behind the radon removal system, respectively. Depending on its position in the system, a radon source is referred to as type 1 or type 2

In order to model the radon concentration inside the detector, we assume a constant $^{222}$Rn production rate $k_1$ by means of emanation from type 1 sources. The total number of $^{222}$Rn atoms inside the detector, $N(t)$, is continuously reduced due to radioactive decay ($t_{1/2} = 3.8$ days) and is also removed by the purification flow. The latter gives rise to an effective exchange time $t_{\text{exchange}} =: 1/f$ with the flow parameter $f$ being the ratio of the purification flow $F_{\text{Xe}}$ in standard liters (sl) per second and the total amount of the xenon inventory $V_{\text{Xe}}$ in sl:

$$f = \frac{F_{\text{Xe}}}{V_{\text{Xe}}}. \tag{1}$$

Additional radon (type 2) is added to the system at a certain rate $k_2$ by the emanation in the gas system in front of the RRS. Inside the RRS the type 2 sources, as well as radon flushed from the detector with a radon particle flux $f \cdot N(t)$, are reduced by a factor $R$, expressing the reduction capability of the RRS. It is defined as the ratio of the radon concentration at the inlet and outlet of the RRS,

$$R \equiv \frac{c_{\text{in}}}{c_{\text{out}}}. \tag{2}$$

The remaining radon atoms re-enter the detector together with additional type 1 sources $k_1$ emanated from parts of the gas system behind the RRS, e.g., the additional pump 2 in the RRS gas interface and the XENON100 detector (see Fig. 2). The change in the number of radon atoms in the detector with time is therefore described by the differential equation

$$\frac{dN(t)}{dt} = k_1 - f \cdot N(t) - \lambda_{\text{Rn}} \cdot N(t) + \frac{k_2 + f \cdot N(t)}{R}, \tag{3}$$

where $f \cdot N(t)$ is again the effective radon particle flux out of the detector and $\lambda_{\text{Rn}} \cdot N(t)$ the decay of $^{222}$Rn using its decay constant $\lambda_{\text{Rn}} = 2.1 \times 10^{-6}\,\text{s}^{-1}$. The term $(k_2 + f \cdot N(t))/R$ describes the number of radon particles that are not removed by the RRS. In this model, we assume the reduction factor $R$ to be constant, i.e., independent on the radon concentration investigated here. Solving this differential equation with the starting condition $N(t = 0) = N_0$, where $N_0$ is the number of radon atoms before starting the removal, we find

$$N(t) = \frac{K}{\Lambda} + \left(N_0 - \frac{K}{\Lambda}\right) \cdot e^{-\Lambda \cdot t};$$
$$\text{with} \quad K = k_1 + \frac{k_2}{R} \quad \text{and} \quad \Lambda = \left(\lambda_{\text{Rn}} + f \cdot \left(1 - \frac{1}{R}\right)\right). \tag{4}$$

For infinitely large times, the above relation simplifies to the equilibrium relation

$$N_{\text{equi}} \stackrel{t \to \infty}{=} \frac{K}{\Lambda} = \frac{k_1 + \frac{k_2}{R}}{\lambda_{\text{Rn}} + f \cdot \left(1 - \frac{1}{R}\right)} \tag{5}$$

$$\stackrel{R \to \infty}{=} \frac{k_1}{\lambda_{\text{Rn}} + f}, \tag{6}$$

where the last step assumes an infinite purification capability $R$. Equation (6) shows that an ideal RRS can fully remove the type 2 sources. The reduction of type 1 sources is limited by the exchange time $1/f$ of the total xenon inventory and thus by the xenon purification flow. This limit for an ideal RRS points to the importance of a high recirculation flow to remove radon from all emanation source types. Using Eq. (5), the radon reduction $r$ inside the XENON100 detector for a given $R$ of the RRS can be defined as

$$r \equiv \frac{N_{\text{equi}}(R = 1)}{N_{\text{equi}}(R)}. \tag{7}$$

## 4 Data analysis

In order to measure the radon reduction due to cryogenic distillation, we employ the XENON100 detector as a sensitive $^{222}$Rn monitor. The $\alpha$-decays of $^{222}$Rn (5.5 MeV) and its direct daughter isotope $^{218}$Po (6.0 MeV) create a clear signal in the primary scintillation light (S1) detected by the photomultiplier tubes (PMTs) of the XENON100 detector [9,14].

Due to PMT saturation effects, an S1 correction map, developed for high energy events in [14,15], is applied to the data to remove light collection inhomogeneities. We require a minimum size for the secondary scintillation signals (S2) of the selected interactions (>75% of $^{222}$Rn S2 signals). This cut removes events suffering from incomplete charge collection (e.g., surface events at the detector's wall). The $\alpha$-events from $^{222}$Rn and $^{218}$Po are clearly identified via their energies derived from the S1 signals (see Fig. 3). Furthermore, we observe the signatures of $^{220}$Rn and its direct daughter $^{216}$Po. The constant emanation of $^{220}$Rn in XENON100 has been investigated in [14,15], but was found to be negligible in this work. In our analysis, solely $^{222}$Rn events are used to monitor the radon concentration in XENON100. These are selected by applying an S1 cut, as shown in Fig. 3, which covers 96% of all $^{222}$Rn events. The overlap with the neighboring $^{218}$Po peak is determined by a fit using crystal ball functions [16]. 8% of the selected $^{222}$Rn events are misidentified $^{218}$Po decays. This is taken into account when determining the $^{222}$Rn activity concentration. Being the direct daughter isotope with a half-life of 3.1 min, $^{218}$Po follows the radon evolution. Consequently, the misidentified events do not impact the investigated reduction factor.

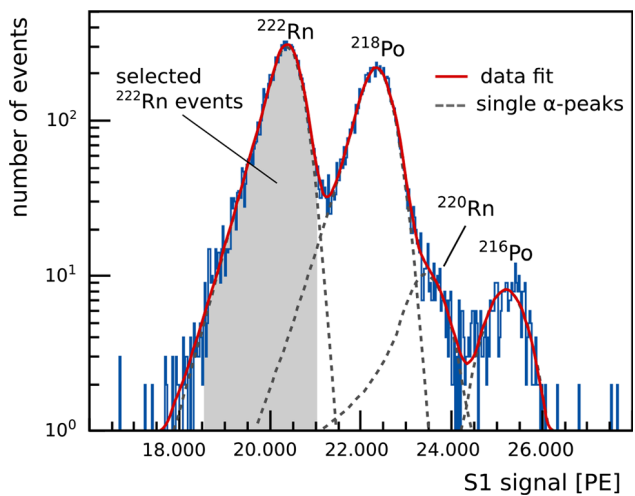

**Fig. 3** $^{222}$Rn decays are selected in XENON100 data by a cut applied to the primary scintillation light signal S1 (*shaded region*)

The subsequent decays of $^{214}$Bi and $^{214}$Po within the $^{222}$Rn decay chain provide an alternative strategy to monitor the radon activity concentration in the detector's liquid xenon target. These so-called BiPo events are clearly identified in a time coincidence analysis, e.g., [14]. The BiPo detection efficiency is reduced by about 50% with respect to the α-analysis described above. This is explained by the limited length of the XENON100 data acquisition time window (400 µs), but also due to ion drift and plate-out effects of $^{222}$Rn progenies [14]. The results of Sect. 5 are therefore based on α-counting while we use the BiPo analysis as a cross-check.

## 5 Radon removal in XENON100

The evolution of the $^{222}$Rn activity concentration in XENON100, acquired during the radon distillation campaign, is shown in Fig. 4. The flat ratio of the activity concentrations obtained by α-analysis and BiPo-counting, respectively, demonstrates consistency of both analyses. According to the operational modes of the gas purification loop, we distinguish seven phases:

I  Pre-distillation phase: the XENON100 detector and its purification loop were operated in their standard (background) mode. The averaged $^{222}$Rn activity concentration is $(33.4 \pm 1.3)$ µBq/kg.
II  Replacement of pump 1: the gas transfer pump 1 was exchanged at the end of phase II. In this phase, the gas purification was stopped several times.
III  First distillation run: xenon gas was looped through the distillation column but by-passed the auxiliary radon emanation source. The observed $^{222}$Rn activity concentration decreased to $(23.2 \pm 1.7)$ µBq/kg.
IV  Standard operation: after the first distillation run, most of the column's radon-enriched LXe reservoir was transferred into the XENON100 detector (abrupt rise of the activity concentration on Dec 23). Residual, radon-enriched xenon stayed inside the column. The $^{222}$Rn activity concentration in XENON100 increased to the new equilibrium value of $(45.4 \pm 1.4)$ µBq/kg. The increased level with respect to phase I is explained by additional $^{222}$Rn sources after the pump exchange. The detector was operated in its standard (background) mode during this phase.
V  Auxiliary radon emanation source opened: xenon gas was circulated through the auxiliary $^{222}$Rn source but bypassing the distillation column. The increased activity concentration was $(518 \pm 8)$ µBq/kg. Since the auxiliary source emanates radon with a constant rate, this level remained constant (emanation equilibrium).
VI  Second distillation run: restart of the online radon removal with the distillation column. The auxiliary emanation source was kept open during this phase. Due to the radon distillation, the $^{222}$Rn activity concentration decreased to $(23.1 \pm 0.7)$ µBq/kg, below the equilibrium value of phase IV. Without radon removal, we would expect to see a constant level as indicated by the dashed, gray line.
VII  Auxiliary radon emanation source closed; radon distillation continued. The activity concentration stayed constant even though the total amount of radon decreased due to decay (dashed, gray line). As we will discuss in the next section, the constant level is an indication that the distillation column removed all radon emanated by the auxiliary emanation source.

In order to quantify the $^{222}$Rn reduction during the first distillation run, we compare the activity concentration measured in phase III with the equilibrium value obtained for phase IV and find a reduction of about a factor $r = 2$. Since in phase IV the system was operated in its standard purification mode, i.e., the XENON100 gas system only, additional emanation sources within the RRS gas interface did not contribute. The observed reduction factor during the first distillation run is therefore interpreted as a lower limit.

In the second distillation run, the $^{222}$Rn activity concentration was reduced by a factor of $r = (22.4 \pm 0.8)$. The higher purification power in the second distillation run is explained by the different composition of the total radon emanation rate with respect to type 1 and type 2 emanation sources. Earlier measurements of the $^{222}$Rn emanation rate of the cryostat and the XENON100 gas system indicated that type 1 sources dominate during the first distillation run [10]. This was different in the second run where the auxiliary emanation source, a type 2 source, dominated the measured $^{222}$Rn activity concentration. As discussed in Sect. 3, our RRS is more efficient

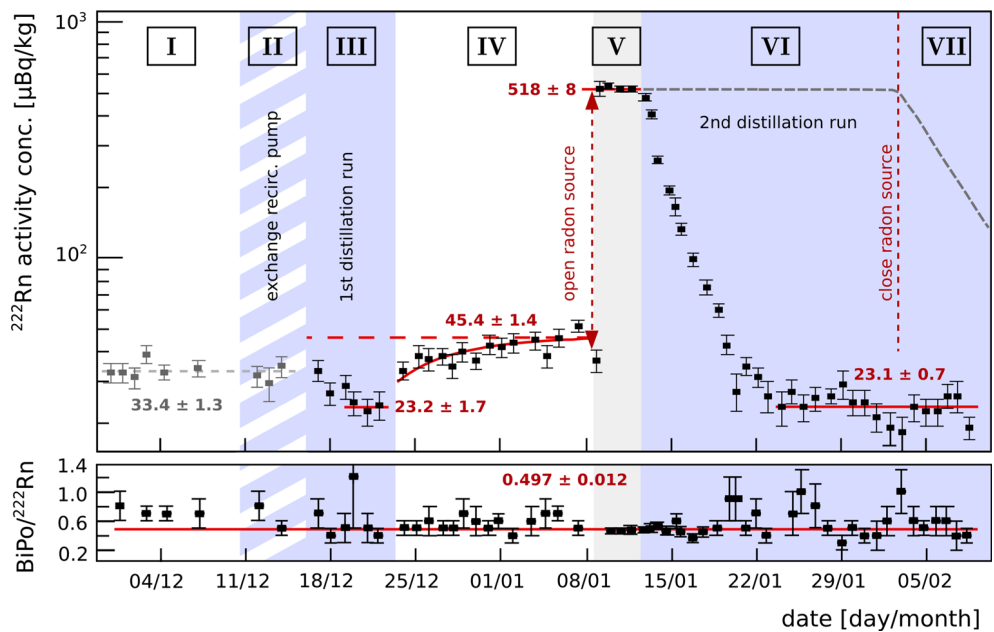

**Fig. 4** *Top panel* Evolution of the $^{222}$Rn activity concentration in XENON100 during the radon distillation campaign. The seven different phases (*roman numerals*) show different operational modes of the detector. The *gray dashed line* shows the expected $^{222}$Rn concentration in the absence of the RRS. *Bottom panel* Comparison of the $^{222}$Rn activity as determined by the $\alpha$-counting method and the BiPo analysis

for type 2 sources while the reduction of type 1 sources is suppressed by the limited purification flow of $F_{Xe} = 4.5$ slpm.

## 6 Reduction capability of the distillation column

In this section, we fit our data using the rate equation model developed in Sect. 3. Doing so, we investigate the radon removal capability of the distillation column, expressed by the reduction factor $R$ (defined in Eq. (2)). We emphasize that due to the limited gas flow, this reduction factor is not equivalent to the observed radon reduction $r$ (defined in Eq. (7)) inside the detector investigated in the previous section. For this analysis we focus on the data from phases V–VII which were acquired under stable conditions.

The rate equation model is based on the assumption of a homogeneous radon distribution inside the detector at any time. We justify this assumption using data acquired when opening the auxiliary radon source at the beginning of phase V. After about two hours, a higher equilibrium in the $\alpha$-rate, originating from $^{222}$Rn and $^{218}$Po, is observed (Fig. 5, bottom). At any time, $^{222}$Rn events are distributed homogeneously inside the sensitive volume. No concentration gradients are found in the Z-position (Fig. 5, top) nor in the radial position.

Since the auxiliary emanation source is of type 2 and can be switched off (see phase VII), we treat it separately in Eq. (4). We define

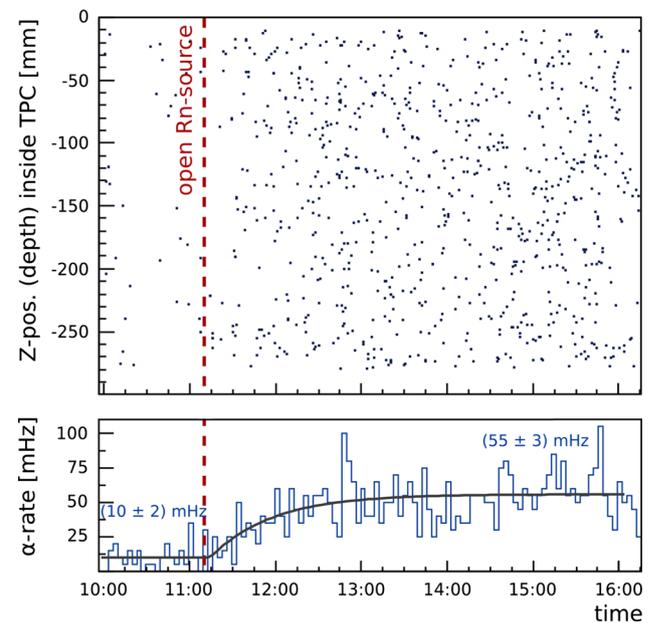

**Fig. 5** Combined $^{222}$Rn and $^{218}$Po $\alpha$-decays before and after opening the auxiliary radon emanation source. The $\alpha$-rate saturates about 2 h after opening the source (*bottom*). The spatial distribution of the events is homogeneous at any time (*top*)

$$K_s = k_1 + \frac{k_2 + k_s}{R} \quad \text{and} \quad K = k_1 + \frac{k_2}{R}, \tag{8}$$

where $k_s$ refers to the emanation of the auxiliary source while $k_2$ describes all other type 2 sources within the system. In order to fit the model to the data, we solve Eq. (3) for the corresponding starting condition for each operational phase. The number of $^{222}$Rn atoms for phases V–VII is given by

$$N_V(t) = (k_1 + k_2 + k_s) \cdot \lambda_{Rn}^{-1} = \text{const.}, \quad (9a)$$

$$N_{VI}(t) = \frac{K_s}{\Lambda} + \left(N_V(t_{VI}) - \frac{K_s}{\Lambda}\right) \cdot e^{-\Lambda \cdot (t - t_{VI})}, \quad (9b)$$

$$N_{VII}(t) = \frac{K}{\Lambda} + \left(N_{VI}(t_{VII}) - \frac{K}{\Lambda}\right) \cdot e^{-\Lambda \cdot (t - t_{VII})}, \quad (9c)$$

where $t_{VI}$ and $t_{VII}$ are the starting times of phases VI and VII, respectively. The factor $\Lambda$ has been defined in Eq. (4). In the following, we use the corresponding $^{222}$Rn activity concentrations ($c_1$, $c_2$ and $c_s$) instead of the rates $k_1$, $k_2$ and $k_s$. This allows for better comparison with the data shown in Fig. 4.

**Table 1** Parameters obtained from fitting the rate equation model, Eq. (9) (a–c), to phases V–VII

| | |
|---|---|
| $c_1$ (type 1) | $(53 \pm 4)\,\mu\text{Bq/kg}$ |
| $c_2$ (type 2) | $(23 \pm 6)\,\mu\text{Bq/kg}$ |
| $c_s$ (aux. em. source) | $(442 \pm 10)\,\mu\text{Bq/kg}$ |
| $f$ (flow parameter) | $(3.1 \pm 0.2) \times 10^{-6}\,\text{s}^{-1}$ |
| R (RRS reduction factor) | $>27$ (95% C.L.) |
| $\chi^2$/ndf | 37.73/28 |

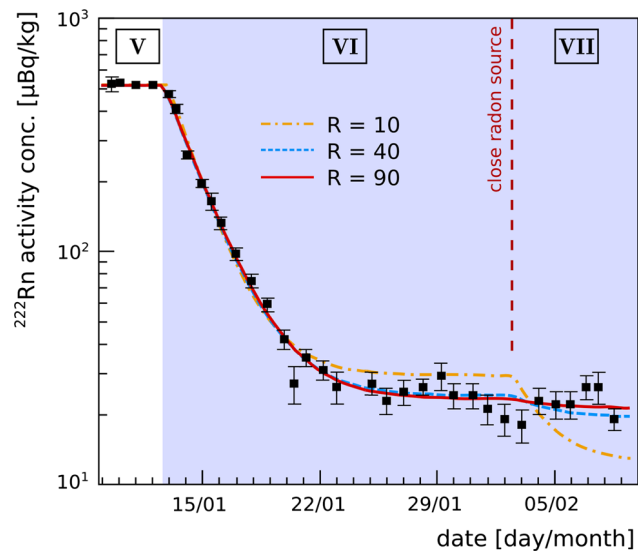

**Fig. 6** $^{222}$Rn activity fit to phases V–VII after constraining the auxiliary emanation source $c_s$ (*red line*). For comparison, the fit is repeated while keeping the reduction factor $R$ fixed at 10, 40 and 90

The fit results, obtained for phases V–VII, are given in Table 1 and Fig. 6 (red curve). The radon contribution from type 1 sources is determined to be $c_1 = (53 \pm 4)\,\mu\text{Bq/kg}$. This is a higher value than the $^{222}$Rn activity concentration measured during phase IV, before the second distillation run. We conclude that the RRS gas interface houses additional type 1 sources, e.g., pump 2 after the distillation column (see Fig. 1).

For this fit, we used $c_s > 400\,\mu\text{Bq/kg}$ as a conservative lower limit for the strength of the auxiliary emanation source. This constraint is based on the $\alpha$-rate monitored when opening the source at the beginning of phase V. As shown in Fig. 5, we can assign about 80% of the increased activity to the auxiliary source. From the fit, we find $c_s = (442 \pm 10)\,\mu\text{Bq/kg}$. The residual type 2 emanation sources are determined to be $c_2 = (23 \pm 6)\,\mu\text{Bq/kg}$. For the flow parameter we obtain $f = (3.1 \pm 0.2) \times 10^{-6}\,\text{s}^{-1}$. We can translate the result for $f$ into a gas purification flow of $F_{Xe} = (5.2 \pm 0.4)$ slpm using the detector's total LXe-mass of $m_{Xe} = (158 \pm 3)$ kg during the distillation campaign. This is a slightly higher mass flow than the measured value of $F_{Xe} = (4.50 \pm 0.05)$ slpm obtained from the flow meters in the purification loop. From this result we conclude that only $(87 \pm 5)$% of the liquid xenon target are effectively purified, assuming a homogeneous radon distribution inside the detector, or that the radon mixes only with this fraction of the total xenon budget.

The best fit value of the distillation column's reduction factor is $R = 91$ (minimum $\chi^2$). Since $R$ is only poorly constrained towards large reduction factors, we quote a lower limit. In a parameter scan the fit is repeated for different, fixed values of $R$. From the $\chi^2$-evolution as a function of $R$, we obtain a lower limit of $R > 27$ at 95% confidence level. For comparison, Fig. 6 shows the fit results obtained when fixing $R$ to 10, 40 and 90, respectively. While $R = 10$ is clearly not supported by our data, the differences are minor for $R > 30$. At larger values of $R$, the significant decrease of the activity concentration after closing the auxiliary emanation source, visible for $R = 10$ in Fig. 6, gets strongly suppressed. A stronger emanation source would be needed in order to probe higher reduction factors.

In Fig. 7, we use the best fit values for $c_1$, $c_2$ and $c_s$ (see Table 1) to predict the radon reduction $r$ inside the XENON100 detector as a function of the purification flow $F_{Xe}$ and the distillation column's reduction power $R$. Assuming that $R$ is independent of the flow, the reduction $r$, defined in Eq. (7), can be calculated for two different scenarios. In the first scenario, the auxiliary emanation source is opened, and thus, the system is dominated by type 2 sources. For a purification flow of $F_{Xe} = 25$ slpm and a given $R = 100$, we expect a reduction $r$ by a factor of 70. In the second scenario, the auxiliary emanation source is closed yielding $c_s = 0\,\mu\text{Bq/kg}$ and thus the system is type 1 dominated as it is the case for the standard (background) operational mode. For a purification flow of $F_{Xe} = 25$ slpm and a given $R = 100$, we expect a reduction $r$ of about 10. In this scenario, the differences between $R = 10$ and $R = 100$ are minor. Consequently, an efficient radon reduction inside the detector can be only achieved by means of sufficiently large purification flows $F_{Xe}$.

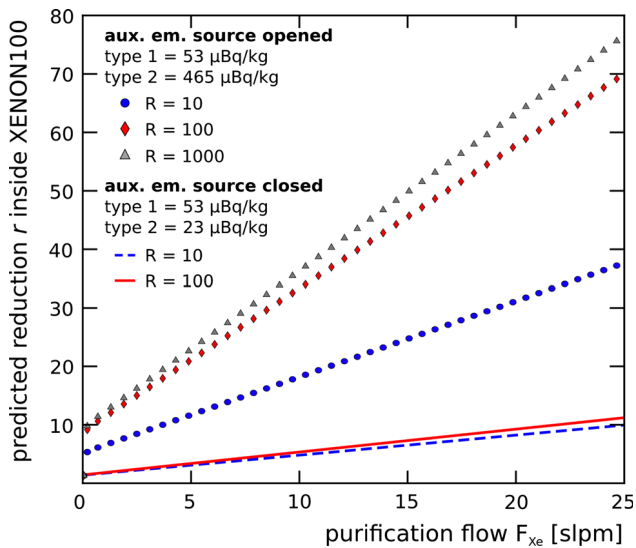

**Fig. 7** Expected $^{222}$Rn reduction $r$ (Eq. (7)) inside the XENON100 detector as a function of the purification flow $F_{Xe}$. We distinguish between the scenarios having the auxiliary emanation source opened (type 2 dominated emanation) and having it closed (type 1 dominated emanation). The reduction $r$ is predicted for different reduction factors $R$ for the radon removal system

## 7 Summary and conclusions

After achieving the dark matter science goal with XENON100, the gas purification system of the detector was extended with a cryogenic distillation column, operated as a $^{222}$Rn removal system. We significantly reduced the $^{222}$Rn activity concentration without any xenon losses (off-gas). The radon reduction capability of the distillation column, defined as the ratio of the $^{222}$Rn concentration in the xenon gas before and after distillation, respectively, was determined to be $R > 27$ (95% C.L.).

These results show the potential of continuous cryogenic distillation as a xenon purification method to reduce $^{222}$Rn induced backgrounds for upcoming liquid xenon detectors. We have shown that the available purification flow is one of the limiting factors for the efficient removal of type 1 $^{222}$Rn emanation sources described in Sect. 3. For upcoming large scale experiments such as XENON1T [4], XENONnT [4] and DARWIN [17], the design and construction of a radon distillation column achieving purification flows up to 200 slpm is currently under investigation.

**Acknowledgements** We gratefully acknowledge support from the National Science Foundation, Swiss National Science Foundation, Deutsche Forschungsgemeinschaft, Max Planck Gesellschaft, German Ministry for Education and Research, Foundation for Fundamental Research on Matter, Weizmann Institute of Science, I-CORE, Initial Training Network Invisibles (Marie Curie Actions, PITNGA-2011-289442), Fundacao para a Ciencia e a Tecnologia, Region des Pays de la Loire, Knut and Alice Wallenberg Foundation, Kavli Foundation, and Istituto Nazionale di Fisica Nucleare. We are grateful to Laboratori Nazionali del Gran Sasso for hosting and supporting the XENON project.